%% file: main.tex
\documentclass[lettersize,journal]{IEEEtran}
\usepackage{amsmath,amsfonts}
\usepackage{algorithmic}
\usepackage{algorithm}
\usepackage{array}
\usepackage[caption=false,font=normalsize,labelfont=sf,textfont=sf]{subfig}
\usepackage{textcomp}
\usepackage{stfloats}
\usepackage{url}
\usepackage{verbatim}
\usepackage{graphicx}
\usepackage{cite}
\usepackage[latin1]{inputenc}
\usepackage[T1]{fontenc}

\usepackage{epstopdf}
\usepackage{booktabs}
\usepackage{multirow}
\graphicspath{{figures/}}
\usepackage{listings}
\usepackage{enumitem}

\usepackage{xcolor}
\usepackage{colortbl}
\definecolor{jade}{rgb}{0.0, 0.66, 0.42}
\definecolor{carolinablue}{rgb}{0.6, 0.73, 0.89}
\definecolor{dkgreen}{rgb}{0,0.6,0}
\definecolor{dkblue}{rgb}{0,0.4,0.5}
\definecolor{gray}{rgb}{0.5,0.5,0.5}
\definecolor{mauve}{rgb}{0.58,0,0.82}

\usepackage[breaklinks,colorlinks,bookmarks=false]{hyperref}
\hypersetup{citecolor=black,linkcolor=black,urlcolor=black}
\usepackage{orcidlink}

\usepackage{flushend}

\usepackage[misc]{ifsym}
\usepackage{bbding}

\usepackage{xspace}
\usepackage{makecell}
\usepackage{multicol}
\usepackage[export]{adjustbox}

\newcommand{\bfit}[1]{\textbf{\textit{#1}}}

\newcommand{\sysname}{ROLoad-PMP\xspace}

\begin{document}

\title{\sysname: Securing Sensitive Operations for Kernels and Bare-Metal Firmware}

\author{Wende Tan\orcidlink{0000-0002-7823-7441}, Chenyang Li\orcidlink{0000-0002-6406-9360}, Yangyu Chen\orcidlink{0009-0007-9891-5610}, Yuan Li\orcidlink{0000-0001-7990-8676},\\Chao Zhang\orcidlink{0000-0001-7894-8828},~\IEEEmembership{Member,~IEEE}, and Jianping Wu,~\IEEEmembership{Fellow,~IEEE}
\thanks{Manuscript created December 2023 and revised July 2024.
Wende Tan and Jianping Wu are with the Department of Computer Science and Technology, Tsinghua University, Beijing, 100084, China.
E-mail: twd2.me@gmail.com.
Chenyang Li is with the Institute of Computer Science and Technology, Peking University, Beijing, 100871, China.
Yangyu Chen is with the College of Computer Science, Chongqing University, Chongqing, 400044, China.
Yuan Li is with Zhongguancun Laboratory, Beijing, 100094, China.
E-mail: lydorazoe@gmail.com.
Chao Zhang is with the Institute for Network Science and Cyberspace, Tsinghua University, Beijing, 100084, China.
E-mail: chaoz@tsinghua.edu.cn.
Chao Zhang and Yuan Li are the corresponding authors.}}

\markboth{IEEE Transactions on Computers}
{Tan \MakeLowercase{\textit{et al.}}: \sysname: Securing Sensitive Operations for Kernels and Bare-Metal Firmware}

\IEEEpubid{0000--0000/00\$00.00~\copyright~2023 IEEE}

\maketitle

\newcommand{\hwoverheads}{$<1.40\%$\xspace}
\newcommand{\overheads}{$<0.853\%$\xspace}

\input{abstract}


\input{intro}
\input{background}
\input{design}
\input{impl}
\input{app}
\input{eval}
\input{discuss}
\input{related}
\input{concl}

\bibliographystyle{IEEEtran}
\bibliography{bib}

\end{document}

%% file: abstract.tex
\begin{abstract}
A common way for attackers to compromise victim systems is hijacking sensitive operations (e.g., control-flow transfers) with attacker-controlled inputs.
Existing solutions in general only protect parts of these targets and have high performance overheads, which are impractical and hard to deploy on systems with limited resources (e.g., IoT devices) or for low-level software like kernels and bare-metal firmware.
In this paper, we present a lightweight hardware-software co-design solution \sysname to protect sensitive operations from being hijacked for low-level software.
First, we propose new instructions, which only load data from read-only memory regions with specific keys, to guarantee the integrity of pointees pointed by (potentially corrupted) data pointers.
Then, we provide a program hardening mechanism to protect sensitive operations, by classifying and placing their operands into read-only memory with different keys at compile-time and loading them with \sysname-family instructions at runtime.
We have implemented an FPGA-based prototype of \sysname based on RISC-V, and demonstrated an important defense application, i.e., forward-edge control-flow integrity.
Results showed that \sysname only costs few extra hardware resources (\hwoverheads).
Moreover, it enables many lightweight (e.g., with negligible overheads \overheads) defenses, and provides broader and stronger security guarantees than existing hardware solutions, e.g., ARM BTI and Intel CET.
\end{abstract}

%% file: intro.tex
\section{Introduction}
\label{sec-intro}

\IEEEPARstart{M}{emory} corruption is one of the most prevalent types of vulnerabilities.
They are often exploited to corrupt operands of sensitive operations (denoted as \textit{sensitive sinks}) with attacker-controlled inputs.
Examples of sensitive sinks include operands used in indirect control-flow transfers, sensitive API invocations, allowlist checks, and application-specific operations (e.g., money transfers).
Attackers can then launch various kinds of attacks by corrupting sensitive sinks.
For instance, attackers can corrupt function pointers or return addresses used by indirect control-flow transfers~\cite{coop, carlini2015control}, or corrupt arguments of sensitive functions like \texttt{system()} and \texttt{execve()}, to hijack the control flow of a process and execute arbitrary code.
Attackers can also corrupt sensitive data like user ID, condition flags, or other metadata (e.g., configurations, policies, shadow memory, etc.) used in security checks to bypass deployed defenses or launch data-only attacks~\cite{ispoglou2018block}.

\IEEEpubidadjcol
A straightforward but effective solution to mitigate this threat is taint analysis~\cite{newsome2005dta}.
Such solutions in general track the data flow of \textit{tainted} input data, and report attacks when tainted data is used at sensitive sinks.
However, they all suffer from over-taint and under-taint issues~\cite{chua2019taintinduce}, introduce overwhelming runtime overheads~\cite{decaf}, and thus cannot be deployed in production environments.

Many other defenses have been proposed to mitigate this threat.
The first type of defenses protects data-flow integrity and stops attacks \textit{at source} by confining corruptions to program states (e.g., sensitive data).
For instance, CPI~\cite{cpi} prevents illegal memory writing from corrupting code pointers, software fault isolation solutions~\cite{sfi} isolate vulnerabilities from sensitive sinks, and StackGuard~\cite{cowan1998stackguard} detects corruptions to sensitive data by placing canaries on the stack.
The second type of defenses secures sensitive sinks with \textit{randomization or isolation}, such that attackers cannot find targets.
For instance, Address Space Layout Randomization (ASLR)~\cite{aslr} randomizes the addresses of sensitive data.
The third type of defenses performs security checks \textit{at sinks}, where potentially corrupted data will be used.
For example, control-flow integrity (CFI)~\cite{cfi, ccfir, bincfi, mcfi, pacitup, kcfi} solutions validate targets before indirect control-flow transfers.
However, these solutions in general only protect part of attack targets, or have very high runtime performance overheads as we will discuss in more details in Section~\ref{sec-related}.

In addition to software-based solutions, many hardware features and hardware-software co-design solutions have been proposed to facilitate or enhance mitigations.
For instance, mechanisms like Intel Memory Protection Extensions (MPX)~\cite{intel-mpx} and ARM Memory Tagging Extension (MTE)~\cite{arm-mte} are introduced to prevent memory corruption \textit{at source}.
Further, mechanisms like Intel Memory Protection Keys (MPK)~\cite{intel-sdm, libmpk}, ARM Domain Access Control Register (DACR)~\cite{arm-dacr}, and HDFI~\cite{hdfi} can be used to \textit{isolate} sensitive data, while IMIX~\cite{imix} places sensitive data in a secure region only accessible by special instructions.
Differently, ARM Pointer Authentication (PA)~\cite{arm-pa, pacitup} protects data-flow integrity by detecting corruptions before it is used, i.e., \textit{at sinks}.
Intel Control-flow Enforcement Technology (CET)~\cite{intel-cet} and ARM Branch Target Identification (BTI)~\cite{arm-bti} can validate indirect-transfer targets and provide CFI support.
However, many of these features are complex to implement or have high runtime overheads, and are hard to deploy on low-end devices.

Therefore, an effective, lightweight, low-overhead, portable, and greatly compatible protection scheme is highly desirable.
In this paper, we propose a hardware-software co-design solution \sysname, consisting of a new set of hardware instructions, an optional Supervisor Binary Interface (SBI) implementation, and compiler extensions, to address these challenges and protect the integrity of sensitive sinks.
\sysname provides a lighter form of data-flow integrity, namely \textit{pointee integrity}, which ensures pointees are not corrupted (i.e., loaded from read-only memory tagged with specific keys or types).
Given this security guarantee, attackers can only feed existing read-only (i.e., non-corrupted) data of expected type to sensitive operations, and thus can hardly launch attacks.

We have built a prototype system supporting \sysname on a field-programmable gate array (FPGA).
Specifically, we extended the RISC-V instruction set architecture (ISA) and augmented the processor core in our prototype, a RISC-V Berkeley Out-of-Order Machine (BOOM) core, to support the \sysname-family instructions.
We also slightly modified the SBI implementation, OpenSBI, to set up read-only memory regions and their keys.
Further, we extended the compiler infrastructure LLVM to provide interfaces for programs to utilize this feature.
Finally, we demonstrated a specific application of \sysname, i.e., type-based forward-edge CFI, to show the effectiveness of \sysname.
Note that \sysname is not limited to this defense application.
\textit{We believe any defense solutions adopting allowlist checks can utilize this feature.}

We evaluated the hardware resource cost of the prototype systems supporting \sysname.
Results showed that \sysname takes few extra hardware resources in terms of FPGA area utilization (\hwoverheads).
We also evaluated the performance of the defense application based on \sysname and found that it can provide a type-based CFI comparable to an existing practical solution~\cite{kcfi} but with much lower runtime overheads (\overheads).
This shows that \sysname is practical and suitable for devices with limited resources (e.g., IoT devices).

In summary, we make the following contributions:

\begin{itemize}[leftmargin=.32cm,noitemsep,topsep=2pt]
    \item We propose a novel lightweight hardware-software co-design solution \sysname, which can provide \textit{pointee integrity} guarantees to secure sensitive sinks.
    \item We introduce a new set of instructions which only load data from read-only memory regions tagged with specific keys to provide pointee integrity, and augment a RISC-V BOOM core to support these instructions on the RISC-V ISA.
    \item We build a prototype supporting this feature on an FPGA, and provide interfaces for software to use this feature.
    \item We demonstrate an important specific defense application of this feature, illustrating its usefulness in practice.
    \item We conduct a thorough evaluation of \sysname, showing it is lightweight, effective, and practical.
\end{itemize}

Note that this paper is a much extended version of our previously published paper, ROLoad~\cite{roload}.
The original ROLoad solution relies on page permissions from memory management units (MMUs) and requires operating system kernels to set up page tables for user-space applications correspondingly.
Thus, it is incapable of protecting operating system kernels themselves or bare-metal firmware, which directly runs on hardware.
In this paper, we revise the ROLoad design, implement it, and conduct experiments based on RISC-V Physical Memory Protection (PMP) rather than MMUs, on an out-of-order core.
Then, we demonstrate that we can also use \sysname to protect low-level software like operating system kernels or bare-metal firmware.
As a complementary solution to the original ROLoad solution, \sysname greatly broadens the application scenarios of ROLoad.

%% file: background.tex
\vspace{-0.3cm}
\section{Background}
\label{sec-background}

\subsection{Sensitive Sinks}

In this paper, we denote operands of sensitive operations as \textit{sensitive sinks}.
There are many types of sensitive sinks in programs, including targets of indirect control-flow transfers (i.e., indirect calls, jumps, and returns), arguments of sensitive APIs (e.g., \texttt{system()}, \texttt{printf()}, or \texttt{execve()}), iteration counts of loops, and metadata of security mechanisms (e.g., shadow memory used by sanitizers, control-flow graphs for CFI defenses, access control policy files, and system configurations).
Applications may have application-specific sensitive sinks, e.g., recipient IP addresses of messages or bank accounts of money transfers.
Most of them essentially have allowlists.

A common way for attackers to compromise victim systems is corrupting sensitive sinks with attacker-controlled inputs, and bypass corresponding allowlists.
Once adversaries can corrupt (or taint) sensitive sinks, they may tamper with the control flow to execute arbitrary code, launch data-only attacks, or break the functionality of programs.

\vspace{-0.3cm}
\subsection{Physical Memory Protection (PMP)}
PMP allows the software to configure permissions, e.g., readable, writable, and executable, for physically contiguous physical memory regions.
This feature is widely supported on modern ISAs, e.g., RISC-V PMP and ARM Memory Protection Unit (MPU)~\cite{arm-mpu}.
For instance, on RISC-V, PMP is configured through PMP control and status registers (CSRs), which can only be accessed by the highest-privileged M-mode code like SBI firmware or security monitors.
However, operating system kernels usually run on the less-privileged S-mode.
Thus, an SBI implementation, e.g., OpenSBI, is required to set up PMP regions for S-mode code when necessary.

%% file: design.tex
\vspace{-0.2cm}
\section{Design}

\vspace{-0.3cm}
\subsection{Threat Model}

The assumptions made in our threat model are consistent with prior work in vulnerability mitigation~\cite{cfi, imix, hdfi, pacitup} and are outlined as follows.

\begin{itemize}[leftmargin=.32cm,noitemsep,topsep=0pt]
    \item We assume that one or more memory-corruption vulnerabilities exist in victim programs, allowing adversaries to repeatedly read from or write to arbitrary addresses that are permitted by memory permission settings.
    \item We assume that DEP is widely deployed and the W$\oplus$X permission policy is applied.
    \item We assume that the CSRs of PMP cannot be tampered after programs have initialized them.
    \item As a hardware-software co-design approach, we assume that the hardware is trusted and cannot be tampered.
          Hardware attacks like Rowhammer~\cite{rowhammer} or side-channel attacks~\cite{spectre, meltdown, mdsattacks} are out of the scope of this paper.
\end{itemize}

\vspace{-0.3cm}
\subsection{Intuition}

Most sensitive sinks essentially have allowlists, which consist of immutable data.
Examples include C++ virtual function tables, customized function pointer tables, format strings, and hardcoded configurations.
On one hand, we can place these immutable data in \textit{tamper-proof areas} to protect them from corruption (or taint).
On the other hand, we can block tainted (or corrupted) data from being used at sensitive sinks, i.e., only data from the tamper-proof areas can be used at sensitive sinks, to mitigate memory-corruption attacks.

In some cases, allowlists are not explicitly defined in programs.
For instance, targets of indirect control transfers could be computed at compile-time or runtime, but are not directly specified in programs.
In these cases, we could pre-compute the allowlists at compile-time.

\begin{figure}[t]
    \centering
    \includegraphics[max width=0.86\columnwidth]{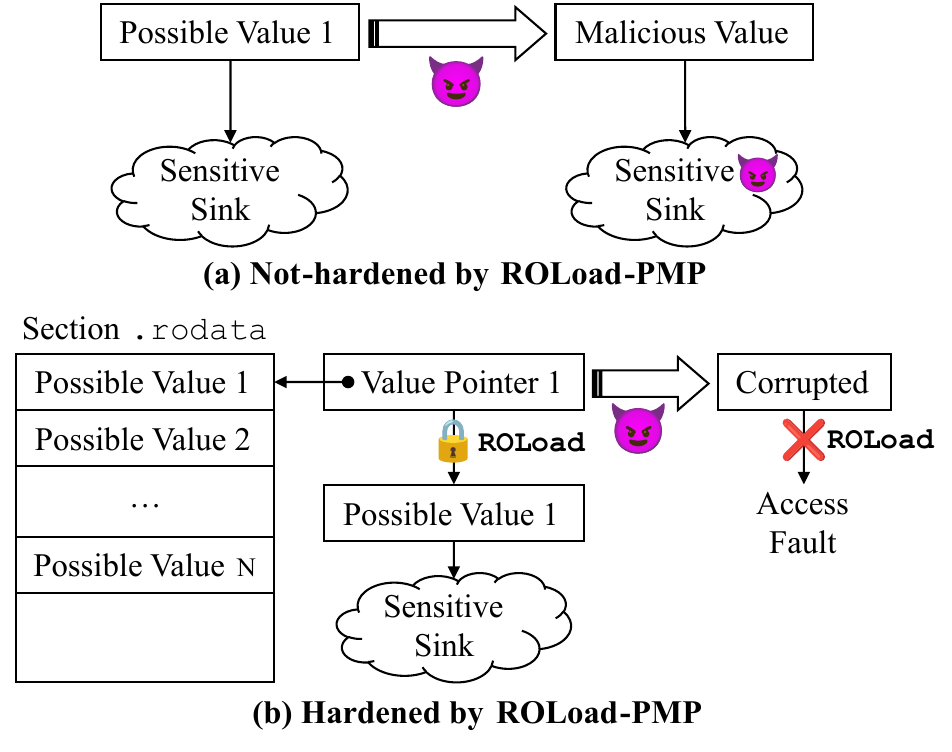}
    \vspace{-0.3cm}
    \caption{Intuition of \sysname's Design.}
    \label{fig-intuition}
    \vspace{-0.5cm}
\end{figure}

Moreover, different types of allowlists could be placed in different tamper-proof areas.
For a given sensitive sink, only a specific allowlist is permitted, i.e., only a specific tamper-proof area can be used.
In this way, attackers can only feed values in the specific type of allowlists to sensitive sinks, and can hardly launch attacks.
Figure~\ref{fig-intuition} demonstrates this intuition.

\vspace{-0.3cm}
\subsection{Design Choices}
On real modern computer systems with physical memory protection (e.g., RISC-V PMP and ARM MPU~\cite{arm-mpu}) or with paging (e.g., RISC-V, ARM, x86, etc.), intuitively, we can choose read-only memory regions (pages) as the tamper-proof areas to place allowlists.
Different types of allowlists may be placed in different read-only regions which are tagged with different \textit{keys} to differentiate from each other.

At runtime, a sensitive sink can only use data loaded from a read-only memory region tagged with a specific key, which is untainted and of the expected type.
In other words, for sensitive sinks, we guarantee the type and integrity of the loaded data (i.e., pointee), rather than intermediate values (e.g., pointers to data and pointers to pointers).

To do so, we introduce a set of simple and lightweight \sysname-family instructions:

\begin{quote}
\sysname-family instructions are a new set of \textit{load} instructions.
Unlike existing load instructions, the memory accessed by these new instructions is restricted to read-only regions with specific keys (tags) to guarantee pointee integrity.
Violating this, i.e., a \sysname-family instruction attempting to access a non-read-only region or a region with undesired key, will cause an access fault.
\end{quote}

The \sysname design is analogous and complementary to DEP (W$\oplus$X), since DEP requires that only memory with execution permissions can be executed, while \sysname requires that only memory with read-only permissions can be used at sensitive sinks.
DEP focuses on the code, but \sysname focuses on the data.
We believe that both of them are necessary in a secure computer system design.
As we will discuss later, the \sysname-family instructions can be easily implemented, have practical security applications, and only introduce very low runtime overheads.

In the rest of this paper, we mainly focus on systems with physical memory protection.
However, our design of \sysname is not limited to these systems, and the design and implementation on systems with paging mechanisms, i.e., memory management units (MMUs), are very similar and have been demonstrated in our previously published paper~\cite{roload}.
It makes \sysname applicable to a wide range of systems, from low-end IoT devices to high-end servers.

\vspace{-0.4cm}
\subsection{The \sysname Solution}
\label{sec-design-sys}

\begin{figure}[t]
    \centering
    \includegraphics[max width=\columnwidth]{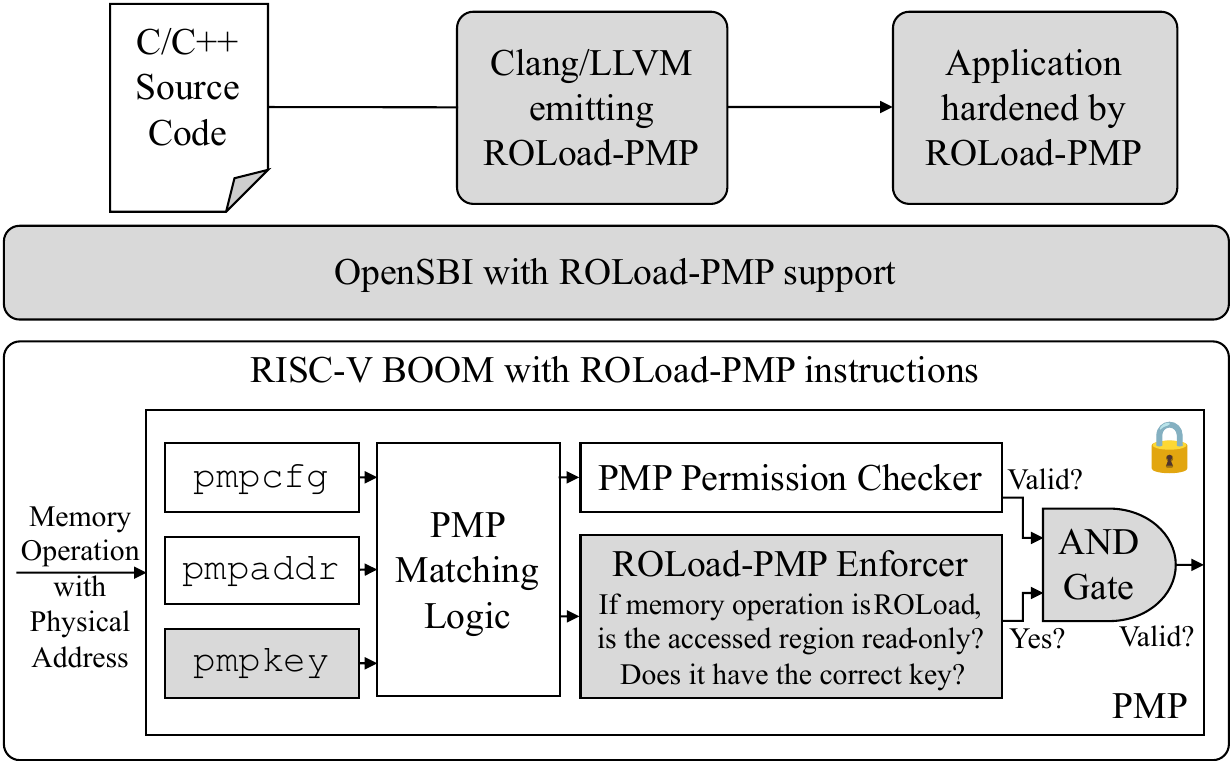}
    \vspace{-0.5cm}
    \caption{Overview of \sysname's Design.
             Components in grey are added (\texttt{pmpkey}, \sysname Enforcer, and the AND gate) or augmented (OpenSBI, Clang/LLVM, and the hardened application).}
    \label{fig-design}
    \vspace{-0.6cm}
\end{figure}

Figure~\ref{fig-design} shows the overview of \sysname.
It consists of the following three major components:
\begin{itemize}[leftmargin=.32cm,noitemsep,topsep=0pt]
    \item the processor core (e.g., a RISC-V BOOM core) at the bottom, responsible for supporting the \sysname-family instructions, which will check the read-only permissions and the keys of pointees automatically.
    \item an optional SBI implementation (e.g., OpenSBI) in the middle, responsible for setting up the read-only memory regions and their keys for S-mode programs.
    M-mode programs can directly configure the CSRs of PMP for themselves and thus do not need this component.
    \item the compiler infrastructure (e.g., Clang/LLVM) at the top of the system, responsible for hardening applications with the proposed \sysname-family instructions.
\end{itemize}

\subsubsection{\sysname-family instruction support}
\sysname-family instructions only load data from read-only memory regions with specific keys.
This permission check and key check are enforced by the processor core efficiently, providing strong security guarantees with low runtime overheads.

Specifically, regardless of ISA, the PMP logic inside a processor can be conceptually divided into two parts, one for PMP matching and the other for PMP permission control.
The former takes the CSRs of PMP and the memory operation as input, and outputs the matching PMP region.
The latter checks whether the memory operation is legal for the matching region.
If it is illegal, the processor generates an access fault.

We logically augment the design of both parts.
First, we augment the former part to support a newly introduced field, namely \textit{key}, for each PMP region.
In this way, each region is associated with a key, and different regions may share a same key.
These keys can be used to represent the kinds, types, or classes of the data stored in these regions, and the actual meanings of the keys are defined by security applications.

Then, \sysname-family instructions will use these keys to ensure that the accessed region is the right one by comparing the keys of these instructions with the keys of the accessed regions.
Specifically, for the latter part, we add a light extra logic to check whether the accessed region is read-only and whether its key is the same as the key of the requesting instruction when \sysname-family instructions are executed.
If yes, then the \sysname-family instructions behave in the same way as normal load instructions.
Otherwise, the processor generates an access fault.
The output of this logic is then ANDed with the original output of the permission control logic.
Thus, the conventional permission check and the newly introduced \sysname checks are done in parallel.

\subsubsection{SBI Interface}
Since S-mode programs, e.g., operating system kernels, cannot access the CSRs of PMP, we provide an optional SBI interface in M-mode responsible for setting up the read-only memory regions and their keys for these S-mode programs, by helping them configure the CSRs of PMP.
M-mode programs can directly configure the CSRs of PMP for themselves and thus do not need this component.

\subsubsection{Compiler Support}
We extend the compiler to provide support for applications that want to utilize \sysname.

The compiler first determines \textit{where sensitive sinks are} and \textit{what their allowlists are}, and then places these allowlists into different read-only regions (e.g., sections in executable files) with different keys (i.e., different tamper-proof areas).

Then, it slightly modifies target programs by replacing regular memory load instructions with \sysname-family instructions at each sensitive sink.
In addition, each \sysname-family instruction is limited to access memory regions with a special key bound to the sensitive sink.

It is worth noting that, this solution is widely applicable to low-level software like bare-metal applications or operating system kernels.

%% file: impl.tex
\vspace{-0.4cm}
\section{Implementation}
\label{sec-impl}

In this section, we dig into more details about implementing the \sysname solution.
For simplicity, we choose to extend the RISC-V ISA to support \sysname-family instructions, and implement a prototype based on an FPGA.
The prototype consists of three major components and the number of lines of code of each component is listed in Table~\ref{tab-lines}.

\begin{table}[t]
    \centering
    \caption{\#Lines of code of each \sysname component.}
    \label{tab-lines}
    \begin{tabular}{ccccc}
      \toprule
      \multirow{2}*{\textbf{Components}} & \multirow{2}*{\textbf{Language}} & \multicolumn{3}{c}{\textbf{\#Lines of Code}} \\
      ~ & ~ & Added & Modified & Total \\\midrule
      RISC-V BOOM & Chisel & 62 & 32 & 94 \\
      OpenSBI & C & 263 & 5 & 268 \\
      LLVM Back-end & C++ and TableGen & 233 & 1 & 234 \\\midrule
      Total & -- & 558 & 38 & 596 \\
      \bottomrule
    \end{tabular}
    \vspace{-0.4cm}
\end{table}

Firstly, we augment the processor core, a RISC-V BOOM core, to support \sysname-family instructions on RISC-V ISA (Section~\ref{sec-impl-riscv}).
We also slightly modify the SBI implementation, OpenSBI, to set up the memory regions and their keys for S-mode programs (Section~\ref{sec-impl-sbi}).
Finally, we extend the compiler to generate proper machine code for \sysname-family instructions (Section~\ref{sec-impl-llvm}).

\vspace{-0.2cm}
\subsection{RISC-V Processor}
\label{sec-impl-riscv}
We chose to extend the RISC-V ISA.
The design of \sysname-family instructions is consistent with RISC-V's design philosophy, e.g., simple, easy to use, and with low cost.
It can be integrated into RISC-V ISA with only slight changes.
However, \sysname is not limited to RISC-V.
Instead, it can be applied to any system with PMP or paging support.

\textbf{Instruction Definition.}
Specifically, we extend the mnemonics of regular memory load operations, including \texttt{ld} and \texttt{lw}, and extend them to the corresponding \sysname versions, e.g., \texttt{ld.ro} and \texttt{lw.ro}, respectively.
Among them, the \texttt{ld} operation is the most common one, since most sensitive operations take full-word-length operands like pointers.
Therefore, we optimize the \texttt{ld.ro} operation to reduce the program size when loading 64-bit words on a 64-bit system.
Specifically, we extended the C extension of RISC-V and added the compressed encoding version of \texttt{ld.ro}, namely \texttt{c.ld.ro}, which is the \sysname version of \texttt{c.ld}.

\textbf{Instruction Encodings.}
Instruction encodings for \sysname-family instructions should be chosen in such a way that they (1) are picked from custom instruction slots or reserved instruction slots, and (2) cause few extra decoding logic circuits.
We finally pick an instruction encoding scheme for \texttt{ld.ro}-family instructions and the compressed version \texttt{c.ld.ro} which is similar to the existing \texttt{ld}-family instructions and \texttt{c.ld}.
Specifically, the machine instruction encoding of \texttt{ld.ro} is the same as \texttt{ld} except that the 3rd least significant bit is changed from \texttt{0} to \texttt{1}.
The machine instruction encoding of \texttt{c.ld.ro} is also the same as \texttt{c.ld} except that the three most significant bits are changed from \texttt{011} to \texttt{100}.
At the same time, to encode keys into machine instructions, we change the meaning of the immediate (up to 12 bits) in these instruction encodings from address offsets to desired keys.

\textbf{Instruction Decoding.}
The instruction decoder is one of the most complicated components within a processor core.
After picking the instruction encodings, we augment the classes \texttt{Instructions}, \texttt{RVCDecoder}, \texttt{XDecode}, and \texttt{X64Decode} in the RISC-V BOOM core to generate decoders of the newly introduced \texttt{ld.ro}-family instructions and \texttt{c.ld.ro}.
Once these instructions are decoded, the core will issue load micro-operations with a new type of memory operations, which carries the keys decoded from the instructions.
We also add this new type to \texttt{MemoryOpConstants}, which will later be used by the \texttt{NBDTLB}\footnote{Actually, RISC-V BOOM cores perform PMP checks in the TLB modules.} to perform the PMP region permission and key checks.

\textbf{PMP Region Permission and Key Checking.}
The \texttt{pmpaddr} and \texttt{pmpcfg} CSRs store the address ranges and the corresponding permissions of all PMP regions.
To associate each PMP region with a key, we first introduce a set of new CSRs, namely \texttt{pmpkey}, to store the keys of all regions.
Then, we augment the PMP matching logic, i.e., the \texttt{PMPChecker} class, to pass the key of the matching PMP region along with the ordinary permissions to the \texttt{NBDTLB} for further checks.
Finally, we add the extra read-only check logic explained in Section~\ref{sec-design-sys} into the \texttt{NBDTLB}.
It ensures that \sysname-family instructions only load data from read-only regions, otherwise an access fault is triggered.
In addition, given the keys, the \texttt{NBDTLB} will also check whether the keys of memory operations and the keys of accessed regions match (the key check logic).
This ensures that \sysname-family instructions only load data from regions with correct keys (i.e., the right regions), otherwise an access fault is triggered.

\vspace{-0.2cm}
\subsection{SBI Interface}
\label{sec-impl-sbi}

To provide an SBI interface for S-mode programs to set up the read-only memory regions and their keys, we augment OpenSBI and add a \texttt{set\_pmp\_ro()} function.
This function can be invoked via an SBI call from the S-mode code, set the specified memory region as read-only, and associate the region with the specified key.
Furthermore, this function will first check whether there is already a region with the specified key and if so, the function will refuse to modify the region to keep the region intact once it has been set up.

\vspace{-0.3cm}
\subsection{LLVM Compiler Back-end}
\label{sec-impl-llvm}

A compiler back-end emits low-level machine instructions for programs written in high-level programming languages.
We extend the RISC-V back-end of LLVM compiler infrastructure to translate \sysname-family instructions to machine code and provide interfaces for applications to utilize via LLVM intermediate representation (LLVM IR) instructions.

Following the design pattern of LLVM, we add the descriptions of \texttt{ld.ro}-family instructions into the \textit{td files} of RISC-V target.
This allows the assembler to recognize \texttt{ld.ro}-family instructions and generate correct machine code.

\textbf{Metadata Interface.}
The interfaces are a new type of metadata, namely \texttt{\sysname-md} metadata.
Users (e.g., defense solutions) associate LLVM IR load instructions of interest with this metadata to indicate that these IR load instructions need to be further protected by \sysname-family instructions.
Keys that will be encoded into \sysname-family instructions are stored in the metadata as well.
We also associate this type of metadata with each machine memory operand (\texttt{MachineMemOperand}) when the compiler performs pattern matching on the DAG (Directed Acyclic Graph) to select and generate machine instructions.

\textbf{Instruction Emission.}
To emit \texttt{ld.ro}-family instructions, we write a machine code pass to visit all LLVM machine instructions and replace all \texttt{ld}-family instructions of interest, whose machine memory operands have \texttt{\sysname-md} metadata, with \texttt{ld.ro}-family instructions.
Since \texttt{ld.ro}-family instructions no longer have any address offset encoded in immediates, we may also insert extra \texttt{addi} instructions.

In addition, the \texttt{c.ld.ro} instruction, i.e., the compressed version,
is automatically emitted by the LLVM assembler when conditions are met, just like the existing compressed versions of regular load instructions, as long as the instruction descriptions of \texttt{c.ld.ro} have been added.

%% file: app.tex
\vspace{-0.2cm}
\section{Applications}
\label{sec-app}
Programs can utilize the \sysname-family instructions to protect the integrity of sensitive sinks, i.e., operands used in sensitive operations.
In this section, we demonstrate an important specific defense application of \sysname, which protects  general forward-edge control-flow transfers, and then discuss other potential applications.

\begin{figure*}[t]
\scriptsize
\begin{minipage}[t]{.30\linewidth}
\begin{lstlisting}[language=c,caption={Pseudo code of two indirect call examples.},label={lst-icall}]
typedef void ((***)func1_t)(...);
typedef int ((***)func2_t)(...);
func1_t func1;
func2_t func2;
// ...
func1 = foo;
// ...
func2 = bar;
// ...
func1();
func2();
\end{lstlisting}
\end{minipage}
\hspace{0.5cm}
\begin{minipage}[t]{0.32\linewidth}
\begin{lstlisting}[language=c,caption={Assembly code of initializing 2 function pointers for \sysname. \texttt{gfpt} refers to global function pointer tables.},label={lst-icall-asm1}]
(*\textcolor{red}{-lui\ \ a0, 0x11}*)
(*\textcolor{red}{-addi a0, a0, 604\ \ \ \# foo}*)
(*\textcolor{dkgreen}{+lui\ \ a0, 0x67}*)
(*\textcolor{dkgreen}{+addi a0, a0, 8\ \ \ \ \ \# gfpt\_foo}*)
(*\ *)sd(*\ \ \ *)a0, -1608(gp)(*\ *)# func1
(*\ *)...
(*\textcolor{red}{-lui\ \ a0, 0x11}*)
(*\textcolor{red}{-addi a0, a0, 616\ \ \ \# bar}*)
(*\textcolor{dkgreen}{+lui\ \ a0, 0x68\ \ \ \ \ \ \# gfpt\_bar}*)
(*\ *)sd(*\ \ \ *)a0, -1600(gp)(*\ *)# func2
\end{lstlisting}
\end{minipage}
\hspace{0.5cm}
\begin{minipage}[t]{0.3\linewidth}
\begin{lstlisting}[language=c,caption={Assembly code of two indirect calls, hardened by \sysname.},label={lst-icall-asm2}]
(*\ *)ld(*\ \ \ \ *)a0, -1608(gp) # func1
(*\textcolor{dkgreen}{+ld.ro a0, (a0), 1}*)
(*\ *)jalr(*\ \ *)a0
(*\ *)ld(*\ \ \ \ *)a0, -1600(gp) # func2
(*\textcolor{dkgreen}{+ld.ro a0, (a0), 2}*)
(*\ *)jalr(*\ \ *)a0

(*\textcolor{dkgreen}{+.section .rodata.key.1}*)
(*\textcolor{dkgreen}{+gfpt\_foo: .quad foo}*)
(*\textcolor{dkgreen}{+.section .rodata.key.2}*)
(*\textcolor{dkgreen}{+gfpt\_bar: .quad bar}*)
\end{lstlisting}
\end{minipage}
\vspace{-0.8cm}
\end{figure*}

\begin{table}[t]
    \centering
    \caption{\#Lines of code of the \sysname-based defense application.}
    \label{tab-app-lines}
    \begin{tabular}{ccccc}
      \toprule
      \multirow{2}*{\textbf{Components}} & \multirow{2}*{\textbf{Language}} & \multicolumn{3}{c}{\textbf{\#Lines of Code}} \\
      ~ & ~ & Added & Modified & Total \\\midrule
      LLVM Pass & C++ & 558 & 0 & 558 \\
      Linux Kernel & C & 238 & 0 & 238 \\\midrule
      Total & -- & 796 & 0 & 796 \\
      \bottomrule
    \end{tabular}
    \vspace{-0.3cm}
\end{table}

\vspace{-0.3cm}
\subsection{Type-Based Forward-Edge CFI}
\label{sec-app2}

Programs have many general indirect calls, especially the programs with callback functions or dynamic dispatching, e.g., the Linux kernel.
Targets of these indirect control-flow transfers (ICT) are sensitive sinks, too.
In practice, adversaries can exploit vulnerabilities to corrupt these ICT targets and launch control-flow hijacking attacks.

CFI is a well-known defense solution against control-flow hijacking attacks, which in general restricts the ICT targets to a limited set at runtime.
Some CFI solutions~\cite{mcfi, pacitup, kcfi}
provide a type-based CFI policy, which allows each ICT to transfer to \textit{address-taken} functions with matching types.
There are several kinds of ICTs, including returns, indirect calls and jumps.
In our prototype, we only focus on the most common forward-edge control-flow transfers, i.e., indirect calls, and aim at providing a type-based CFI policy.

\subsubsection{Defense Principles}
\sysname can be utilized to enforce the same policy as type-based forward-edge CFI.
Under this policy, the legitimate transfer targets of indirect calls are the entry points of address-taken functions, whose addresses must be taken (i.e., used) in the source code and will keep intact at runtime.
Besides, the function types (i.e., the types of parameters and return values) of legitimate functions should match the expected function types of the indirect calls.

Therefore, we first identify all address-taken functions and their types, and place their addresses into read-only memory regions with keys grouped by the types.
Then, we modify indirect call instructions to only use function addresses loaded from these read-only memory regions with matching keys, rather than using function addresses stored in mutable function pointer variables as before.
In this way, \sysname ensures that indirect calls only transfer to address-taken functions with correct types, providing a type-based forward-edge CFI.

Listings~\ref{lst-icall}, \ref{lst-icall-asm1}, and \ref{lst-icall-asm2} demonstrate an example of this application.
In this example, two global function pointer tables (GFPTs) are created and put in read-only memory regions with the keys of 1 and 2, and are both initialized to an array of addresses of address-taken functions, i.e., \texttt{foo} and \texttt{bar} (Lines 8$\sim$11 in Listing~\ref{lst-icall-asm2}).
Function pointers in the program are replaced with pointers pointing to GFPT items, as shown in Listing~\ref{lst-icall-asm1}.
When function pointers are used, the corresponding items are loaded by \texttt{ld.ro} instructions (Lines 2 and 5 in Listing~\ref{lst-icall-asm2}), ensuring that the function addresses are loaded from the read-only arrays (i.e., GFPTs) with the correct keys of 1 and 2.
Then, these items, which are the actual addresses of the callee functions, are used by the following indirect call instructions (Lines 3 and 6 in Listing~\ref{lst-icall-asm2}).

\subsubsection{Implementation Details}
Function pointers in C or C++ programs in general store addresses of callee functions.
To implement a CFI mechanism utilizing \sysname, we need to make original function pointers point to read-only memory regions with specific keys, where real function pointers are stored.
We therefore write an LLVM optimization pass to analyze LLVM IR and transform instructions that define or use function pointers.
The number of lines of C++ code to implement this application is shown in Table~\ref{tab-app-lines}.

Specifically, we first compute arrays of addresses of functions by iterating over the function list and the function alias list.
For each function or function alias, we hash the string representation of its function type to compute a key, and then put the function address in the corresponding array according to the key.
Functions whose addresses are not used, i.e., non-address-taken functions, are automatically removed from the arrays during linking, so these arrays only hold the addresses of address-taken functions.
These arrays are named \texttt{gfpt}s in the code and are stored as constants into read-only regions with the corresponding keys.

Secondly, we identify all \textit{use points} of function addresses, and augment the program to replace function addresses at these points with pointers pointing to \texttt{gfpt} items.
These use points are usually the \textit{definition points} of function pointers, including (1) initializers of global function pointers, (2) \textit{store} instructions that store addresses into function pointers, (3) \textit{call} instructions that take function addresses as arguments, (4) \textit{return} instructions that return function addresses, and so on.
In other words, these program points \textit{create} (aka. \textit{generate}) function pointers from the function addresses.

Finally, we identify all \textit{use points} of function pointers that need the actual function addresses.
Since we have replaced function pointers with pointers to \texttt{gfpt} items, we have to restore the original (i.e., actual) function addresses for function pointers.
We insert a \sysname-family load instruction (i.e., an LLVM IR load instruction with \texttt{\sysname-md} metadata) with our computed key before each use point to load the actual function addresses from the modified function pointers.
The key is computed from the used function type using the same algorithm.
This \sysname-family load instruction ensures the integrity of function addresses to use, and thus secures these use points.
These use points in general are sensitive sinks.
Some of them are direct \textit{dereferences} of function pointers, which need the actual function addresses immediately, including indirect call instructions.
Other use points may also need the actual function addresses but will dereference function pointers later.
These use points include (1) \textit{call} instructions that call functions located in unprotected libraries and take function pointers as arguments, (2) inline assembly codes that take function pointers as operands, (3) \textit{compare} instructions that compare with other function pointers, and so on.
In addition, some use points do not need the actual function addresses, e.g., a \textit{store} instruction that copies a function pointer to another.
Function pointers at these use points are kept unmodified.

\subsection{Other Application Scenarios}
\label{sec-other-app}
It is worth noting that the aforementioned application reveals a defense principle: the targets allowed in sensitive sinks are in kinds of allowlists, i.e., sets of legitimate functions.
We believe that allowlist checks all can be enhanced by \sysname, and defenses utilizing allowlist checks can thus benefit from it.

Specifically, given an allowlist check, we can first place the allowlists into read-only memory regions tagged with unique keys, and then transform the allowlist check to a \sysname check, i.e., ensuring the targets are loaded from read-only regions tagged with the correct keys.
It thus ensures that the targets are in the allowlists, prohibiting many attacks with low overheads.

For instance, it can be applied to backward control-flow transfers (i.e., return instructions) too, where the allowlists are sets of legitimate return sites.
It can also be applied to operating system kernels in which many structure pointers (e.g., device descriptors, operation structures) have limited value choices, where the sets of available device descriptors or operation structures can be viewed as the allowlists.
Due to the space limitation, we skip the demonstration of these defense applications and leave it as future work.

%% file: eval.tex
\vspace{-0.25cm}
\section{Evaluation}
\label{sec-eval}

\begin{table}[b]
    \centering
    \vspace{-0.4cm}
    \caption{Configurations of our prototype system.}
    \label{tab-hw-config}
    \begin{tabular}{cc}
        \toprule
        \textbf{Components} & \textbf{Configurations} \\\midrule
        \textbf{ISA Extensions} & RV64IMAC with M, S, and U modes \\
        \textbf{BOOM Config} & \texttt{SmallBooms} \\
        \textbf{Caches} & 16 KiB 4-way L1I\$, 16 KiB 4-way L1D\$  \\
        \textbf{TLBs} & 32-entry I-TLB, 8-entry D-TLB \\
        \textbf{PMP} & 16 entries (regions) \\\midrule
        \multirow{2}*{\textbf{Peripherals}} & Xilinx MIG for 4 GiB DDR4 components \\
        ~ & Xilinx AXI Ethernet, 64 KiB Boot ROM \\
        \bottomrule
    \end{tabular}
\end{table}

\begin{table*}[ht]
  \centering
  \caption{Hardware resource cost of systems without and with \sysname when synthesized on an FPGA.
          }
  \label{tab-hw-cost}
  \begin{tabular}{c|cccc|cccccc}
    \toprule
    \multirow{2}*{} & \multicolumn{4}{c|}{\textbf{RISC-V BOOM cores}} & \multicolumn{6}{c}{\textbf{Whole Systems}} \\
    ~ & \#LUT & \% & \#FF & \% & \#LUT & \% & \#FF & \% & WSS (ns) & $F_{\textit{max}}$ (MHz) \\\midrule
    \textbf{without \texttt{ld.ro}} & 57,561 & -- & 38,111 & -- & 82,318 & -- & 73,561 & -- & 0.177 & 101.81 \\
    \textbf{with \texttt{ld.ro}} & 57,882 & $+$0.55766 & 38,641 & $+$1.39067 & 82,622 & $+$0.36929 & 74,103 & $+$0.73680 & 0.188 & 101.91 \\
    \bottomrule
  \end{tabular}
  \vspace{-0.4cm}
\end{table*}

In this section, we evaluate our prototype system featuring \sysname and applications hardened by \sysname-family instructions by answering the following questions:

\begin{itemize}[leftmargin=.32cm,noitemsep,topsep=0pt]
    \item \textbf{RQ1: Lightweight Hardware Implementation:}
          Is it low-cost and lightweight to implement \sysname-family instructions on hardware
          (e.g., on FPGAs)? (Section~\ref{sec-logic-usage})
    \item \textbf{RQ2: Low Runtime Overheads:}
          Does a system with \sysname run as fast as an unmodified system?
          How much execution time overhead do applications hardened by \sysname incur?
          (Section~\ref{sec-app-perf})
    \item \textbf{RQ3: Strong Security Guarantees:}
          What security guarantees do \sysname and applications hardened by \sysname provide?
          (Section~\ref{sec-app-security})
\end{itemize}

\vspace{-0.3cm}
\subsection{Hardware Resource Cost}
\label{sec-logic-usage}

To evaluate the hardware resource cost of \sysname-family instructions, we instantiate the original and the modified RISC-V BOOM cores and synthesize them on an FPGA.
Table~\ref{tab-hw-config} shows the overall configurations of our prototype systems.
Note that we have not instantiated a floating-point unit (FPU), i.e., F and D extensions of RISC-V ISA, to highlight extra hardware resource cost caused by \texttt{ld.ro} instructions.
We then integrate the above RISC-V BOOM cores with necessary peripherals, i.e., a Xilinx Memory Interface Generator, a Xilinx AXI Ethernet Subsystem, and a 64KiB boot ROM, to build the prototype systems.
These systems can boot the Linux operating system from a network.

We synthesize and map our prototype systems to a Xilinx Kintex UltraScale FPGA (\texttt{XCKU060}) on a commodity FPGA development board using Xilinx Vivado 2022.2.
The target frequency of synthesis is $F_{\textit{target}} = 100.00\text{MHz}$, which is approximately the maximum frequency of the RISC-V BOOM cores on our FPGA.
We also synthesize the RISC-V BOOM cores out of context without peripheral to evaluate the hardware resource cost of RISC-V BOOM cores themselves.

The hardware resource cost, the worst setup slack (WSS), and the maximum frequency $F_{\textit{max}}$ are presented in Table~\ref{tab-hw-cost}.
The hardware resource cost is measured in terms of the number of lookup tables (LUTs) and flip-flops (FFs).
The maximum frequency is calculated by $F_{\textit{max}} = \frac{1}{\frac{1}{F_{\textit{target}}} - \textit{WSS}}$.

The results show that the extra hardware resource cost on an FPGA measured in terms of the number of LUTs and FFs are both low (\hwoverheads) when the RISC-V BOOM core with \texttt{ld.ro} instructions is synthesized without peripheral as well as when it is integrated and synthesized in a whole system.
Besides, the maximum frequency of the system supporting \texttt{ld.ro} instructions is approximately the same as that of the system without \texttt{ld.ro} instructions, since our modifications do not affect the critical path of the logic circuits of the processor core.
This gives us evidence that \bfit{\sysname-family instructions can be adopted in real-world systems with low hardware cost.}

\vspace{-0.3cm}
\subsection{Defense Application Evaluation}
As we have mentioned, \sysname can be utilized to secure sensitive sinks and harden programs.
In this section, we evaluate its performance and security guarantees based on the specific application, i.e., type-based forward-edge CFI.

\subsubsection{Performance}
\label{sec-app-perf}

To evaluate the performance impact of the security application mentioned in Section~\ref{sec-app}, we measured the overall performance of the Linux kernel hardened by the application.
To compare, we also measured the Linux kernel hardened by the competitor, i.e., KCFI~\cite{kcfi}.
KCFI is a practical type-based forward-edge CFI solution that has already been integrated into the LLVM compiler suite and can be easily enabled through a compilation flag for the Linux kernel.
Thus, we chose to compare \sysname with KCFI.
However, it would be better to compare \sysname with hardware-assisted solutions, but we did not find such solutions that are as practical as KCFI.
Note that CFI solutions for operating system kernels usually require the page tables of the kernels to be protected~\cite{ptstore}.
Otherwise, these solutions would be bypassed.
In this paper, we only use the security application to demonstrate the practicality of applying \sysname to low-level software and evaluate the performance impact of \sysname.
Thus, protecting page tables is out of the scope of this paper.

We used the Linux kernel 6.6 with \texttt{defconfig} plus necessary device drivers for our prototype system.
We implemented the security application mentioned in Section~\ref{sec-app2} based on LLVM 18 (\texttt{88dd9813}).
According to the hardware configurations shown in Table~\ref{tab-hw-config}, excluding the PMP entries used by OpenSBI, there were 11 PMP regions available.
We used all of them and assigned a unique key for each region.
Totally, we used 11 unique keys.
To utilize \sysname for the kernel, we also augmented the initialization procedure and the linker script of the kernel to set up the read-only memory regions through SBI calls.
The number of lines of code is presented in Table~\ref{tab-app-lines}.
We then conducted all the following experiments on three experimental setups: (1) the system and kernel without any modifications (i.e., the baseline), (2) the system without processor modifications and the kernel hardened by KCFI (i.e., KCFI), and (3) the system with \sysname and the kernel hardened by \sysname (i.e., \sysname).
All benchmarks themselves were unmodified.
Specifically, for the \sysname setup, we did not notice significant time usage increments of the LLVM compilation and the system initialization procedure.
Since they only need to be done once before the system boots, we believe that their influence will be negligible.

\begin{figure}[b]
  \centering
  \vspace{-0.3cm}
  \includegraphics[max width=0.9\columnwidth]{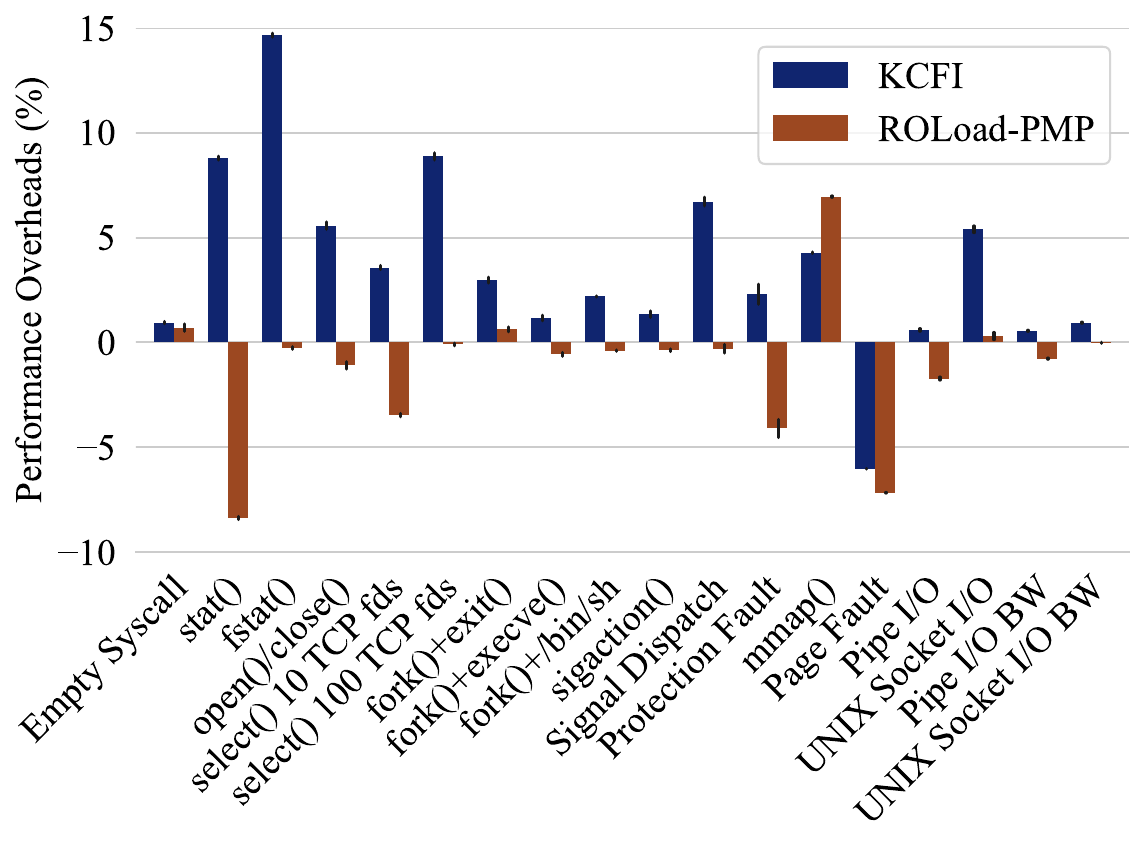}
  \vspace{-0.3cm}
  \caption{Performance overheads of LMBench microbenchmarks on the Linux kernel hardened by \sysname and its competitor KCFI.
  \textit{BW} refers to the bandwidth.}
  \label{fig-lmbench}
\end{figure}

\textbf{Microbenchmarks.}
We first leverage LMBench 3.0-a9 as our microbenchmark suite to evaluate the performance impact on system calls and trap handlers.
We execute each benchmark 1,000 times and report the average relative performance overheads in Figure~\ref{fig-lmbench}.
The results show that, on average, \sysname has a negligible performance impact on system calls and trap handlers.

\begin{figure}[t]
  \centering
  \includegraphics[max width=0.9\columnwidth]{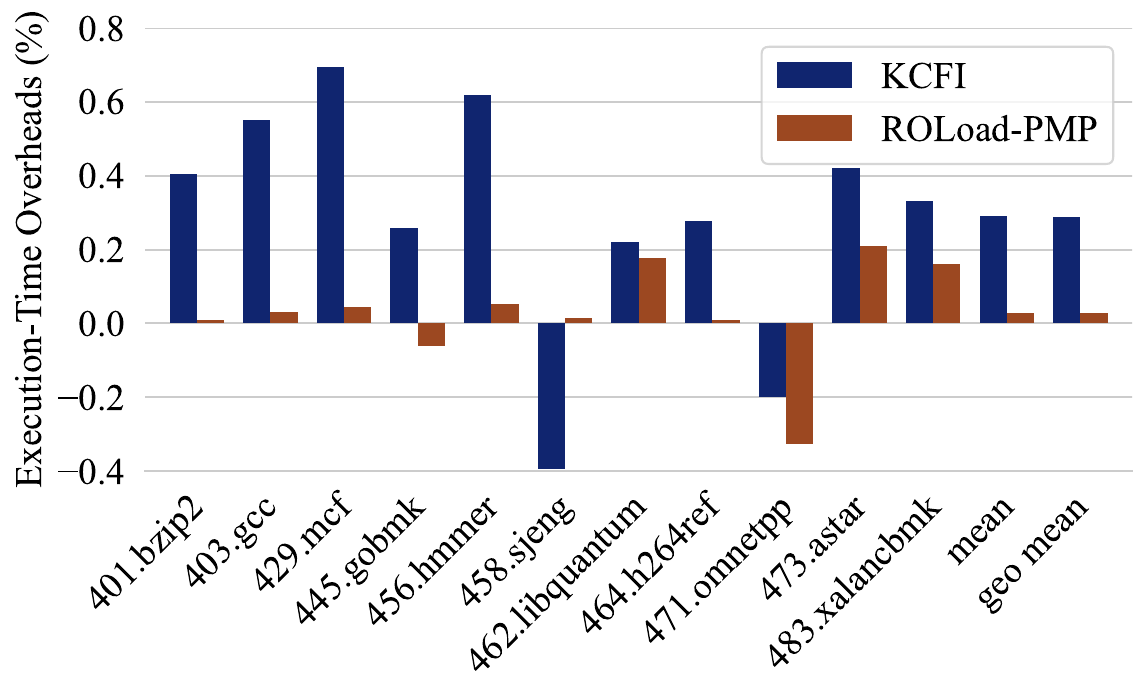}
  \vspace{-0.3cm}
  \caption{Execution-time overheads of SPEC CINT2006 benchmarks on the Linux kernel hardened by \sysname and its competitor KCFI.}
  \label{fig-spec}
  \vspace{-0.2cm}
\end{figure}

\begin{figure}[t]
  \centering
  \includegraphics[max width=0.9\columnwidth]{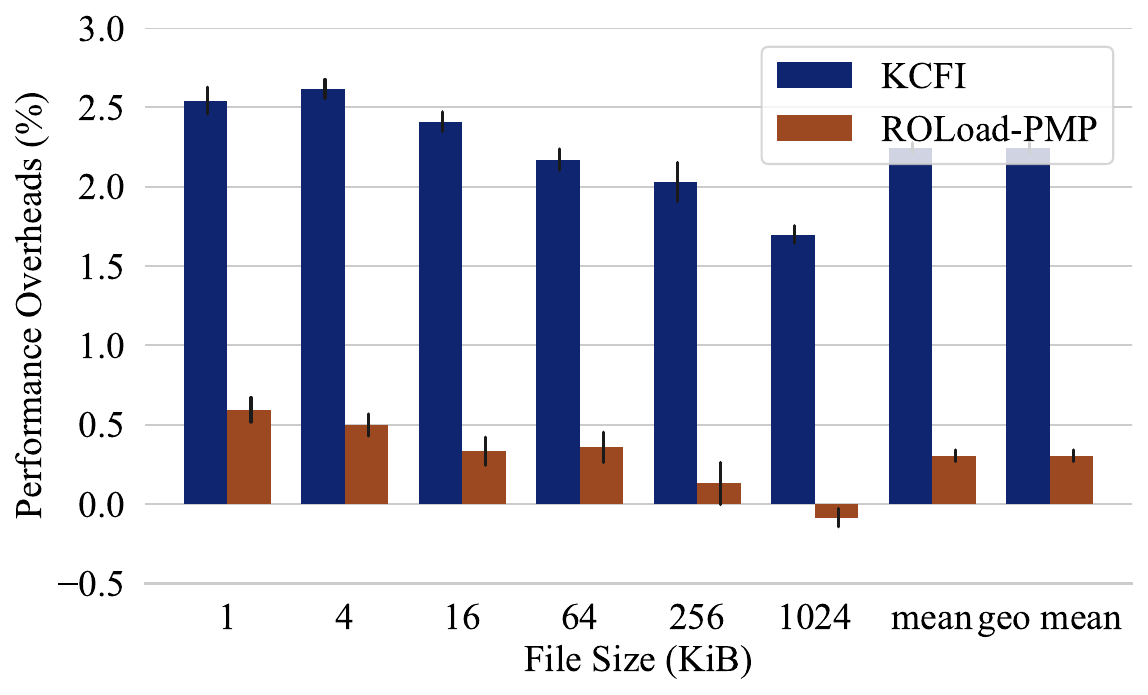}
  \vspace{-0.3cm}
  \caption{Performance overheads of Nginx on the Linux kernel hardened by \sysname and its competitor KCFI.
           10,000 requests in total, 100 concurrent requests.}
  \label{fig-nginx}
  \vspace{-0.3cm}
\end{figure}

\begin{figure}[t]
  \centering
  \includegraphics[max width=0.9\columnwidth]{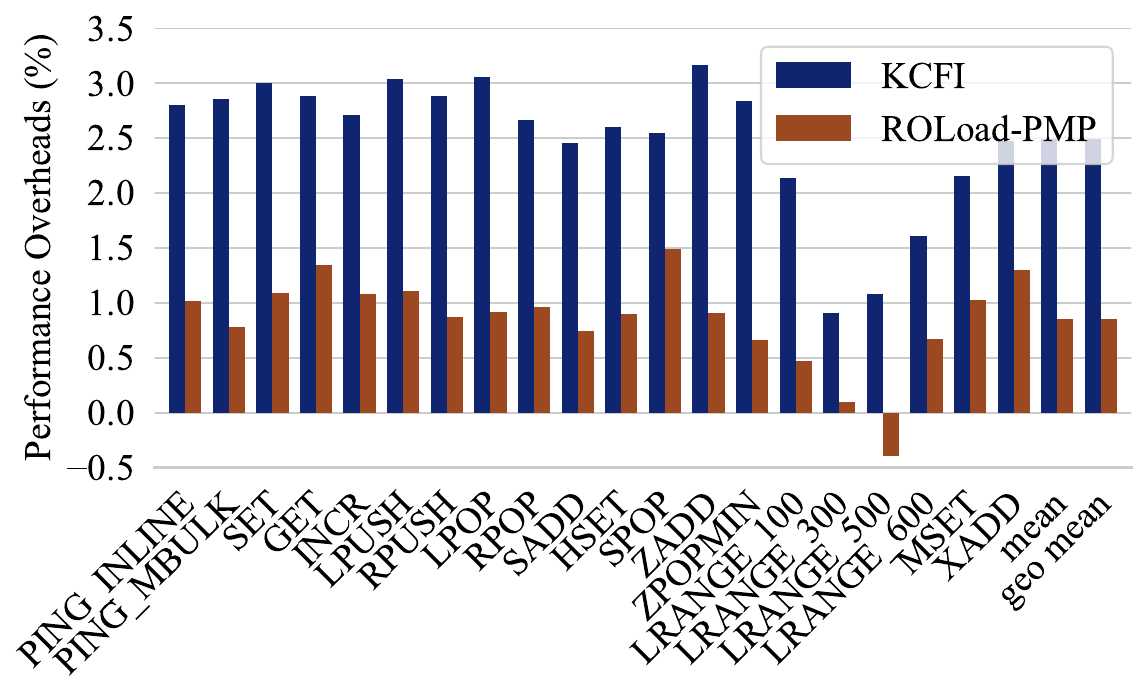}
  \vspace{-0.3cm}
  \caption{Performance overheads of Redis on the Linux kernel hardened by \sysname and its competitor KCFI.
           100,000 requests for each test, 50 parallel connections.}
  \label{fig-redis}
  \vspace{-0.45cm}
\end{figure}

\textbf{Macro Benchmarks.}
To evaluate the overall performance of the system with \sysname and the kernel hardened by \sysname, we also run a CPU-intensive benchmark, i.e., the SPEC CPU2006 benchmark suite, as well as two kernel-intensive benchmarks, i.e., Nginx 1.24.0 benchmark and Redis 7.2.3 benchmark.
Specifically, since we have not enabled the FPU, we use the integral subset of the SPEC CPU2006, i.e., SPEC CINT2006.
Among them, \textit{400.perlbench} is excluded due to compilation failure.
For the Nginx benchmark, we measure its HTTP throughput using 10,000 requests with 100 concurrent requests.
We repeat this experiment 10 times and report the relative performance overheads calculated by using the average throughput.
For the Redis benchmark, we use 100,000 requests for each test with 50 parallel connections.
Then, we compute the relative performance overheads.
Figures~\ref{fig-spec}, \ref{fig-nginx}, and \ref{fig-redis} show the results.

Since all the benchmarks can complete successfully, we can confirm that the kernel hardened by \sysname works well.
The results show that the average execution-time overheads of the CPU-intensive SPEC CINT2006 benchmarks on the hardened kernel are negligible ($<0.030\%$), which are about one order of magnitude lower than its competitor KCFI ($<0.289\%$).
Besides, the results of the kernel-intensive benchmarks show that the average performance overheads of the Nginx and Redis benchmarks are both low ($<0.305\%$ and \overheads, respectively), which are also much lower than its competitor KCFI ($<2.24\%$ and $<2.48\%$, respectively).

We also notice some negative overheads.
As our transformation slightly modified the code and data memory layout of the applications, the TLB and cache hit rates changed.
Thus, we believe that the negative overheads were due to the influence of the changed hit rates overwhelming the actual overheads because the actual overheads were too low.

We can conclude that the defense application that we have demonstrated introduces very low overheads.
It implies that \bfit{defense solutions that utilize \sysname can protect real-world software with very little extra runtime cost}.

\subsubsection{Security}
\label{sec-app-security}

The security guarantees provided by \sysname depend on the underlying hardware primitives, i.e., the \sysname-family instructions, and its applications.
We compare the \sysname-family instructions with others later in Section~\ref{sec-related}.

The security application mentioned in Section~\ref{sec-app} restricts the targets of indirect calls to those with the matching types (i.e., keys), and thus implements a type-based CFI.
It provides security guarantees stronger than coarse-grained CFI solutions (e.g.~\cite{ccfir, bincfi}), which do not take function types into consideration.
Furthermore, the type-based CFI has been extensively studied and its advantages and weaknesses are well-known~\cite{coop, vtint, cfi, kcfi}.
We can confirm that \bfit{our \sysname solution provides comparable security guarantees but with much lower runtime performance overheads}.

%% file: discuss.tex
\vspace{-0.2cm}
\section{Discussion}

In this section, we discuss the advantages and limitations of \sysname and its applications.

\textbf{Pointer Transitivity and Pointee Reuse Attacks.}
It is important to note that like prior lightweight hardware-based solutions, e.g., DEP~\cite{dep}, ARM BTI~\cite{arm-bti}, and Intel CET~\cite{intel-cet}, our \sysname solution may also suffer from pointee reuse attacks as pointees in read-only memory regions with keys may be reused by adversaries~\cite{coop}.
For example, a sophisticated adversary can corrupt pointers to data or pointers to pointers, and so on, to reuse existing data in any read-only regions with matching keys.
These crafted pointers may pass pointee integrity checks enforced by \sysname-family instructions, since they do point to read-only regions with correct keys.
However, we argue that the remaining attack surface is minimal and this attack is much harder to realize in practice, as attackers can only feed values in the specific allowlists to sensitive operations.

In the future, we can lengthen memory region keys when possible or use more levels of indirection to achieve longer equivalent keys under existing \sysname hardware.
Users can balance the key length and security guarantees.

\textbf{Portability.}
Although we implement and demonstrate \sysname on the RISC-V ISA, the design of \sysname-family instructions is architecture-independent.
For ISAs that are also \textit{reduced instruction sets}, including widely-used ARM ISA, we can easily implement \sysname-family instructions on these architectures in the same way that we have done on the RISC-V ISA.
This is left as our future work.
For \textit{complex instruction sets}, especially x86 ISA, almost all arithmetic and logic instructions may load from or store to the memory.
Instead of introducing new instructions, we can introduce a new instruction prefix, which indicates that memory operands of the following instruction have to be loaded from read-only memory regions with specific keys.

\textbf{Limitations.}
As we have mentioned in Section~\ref{sec-other-app}, we believe that \sysname can protect sensitive sinks whose legitimate value sets (i.e., allowlists) can be determined during compilation time and do not change at runtime, with low overheads.
However, \sysname is not suitable for sensitive sinks whose allowlists cannot be determined during compilation time (e.g., session keys generated by key-exchange protocols) since we cannot pre-compute these mutable allowlists and put them into read-only memory.

%% file: related.tex
\vspace{-0.3cm}
\section{Related Work}
\label{sec-related}

In this section, we review some related defense mechanisms and compare \sysname with them.

\textbf{Data Execution Prevention (DEP).}
DEP is a defense method to defeat code injection and code corruption attacks by preventing any (writable) data from being executed~\cite{dep}.
This can be implemented both in hardware by adding a permission bit to page table entries (e.g., AMD NX-bit, Intel XD-bit, ARM XN-bit, and RISC-V X-bit), and in software by emulating.
For modern systems with these permission bits, loaders and operating systems configure page tables with the W$\oplus$X policy to realize DEP.
The basic ideas of DEP and \sysname are very simple and similar, since \sysname requires that sensitive operations are only fed with read-only data with matching keys, while DEP (W$\oplus$X) requires that only read-only \textit{data} can be executed as instructions.

\textbf{Control-Flow Integrity (CFI).}
As we have mentioned, CFI is a well-known defense solution against control-flow hijacking attacks~\cite{cfi, kcfi}, which in general restricts the ICT targets to a limited
set at runtime.

Type-based CFI schemes include not only pure-software-based KCFI~\cite{kcfi}, but also PARTS~\cite{pacitup} utilizing ARM PA, and others~\cite{mcfi}.
The runtime overheads of these CFI mechanisms have been reduced as much as possible, but are still rather high for low-end systems (e.g., IoT devices).
Even for PARTS using hardware feature, the performance overheads are still around $1\%$.
Compared with these mechanisms with the same security strength, the performance overheads of the CFI scheme mentioned in Section~\ref{sec-app2} are almost negligible.
Thus, \sysname is more practical on systems with limited resources.

Several recent hardware features have been proposed to facilitate CFI mechanisms, including Intel CET~\cite{intel-cet} and ARM BTI~\cite{arm-bti}.
Intel CET includes a shadow stack and an indirect branch tracking (IBT) mechanism, and requires to employ an extra architectural state, which needs to be maintained when the operating system kernel is performing context switching or handling interrupts.
Extra architectural states could increase the complexity of the operating system kernel as well as the processor core.
Similarly, ARM BTI also needs some extra architectural states.
Although they are perfect hardware features regardless of the extra architectural states, they only focus on CFI.
Besides, the security strength of Intel IBT and ARM BTI is weaker than type-based CFI solutions, since they only provide coarse-grained CFI.

\begin{figure}[t]
    \centering
    \includegraphics[max width=.85\columnwidth]{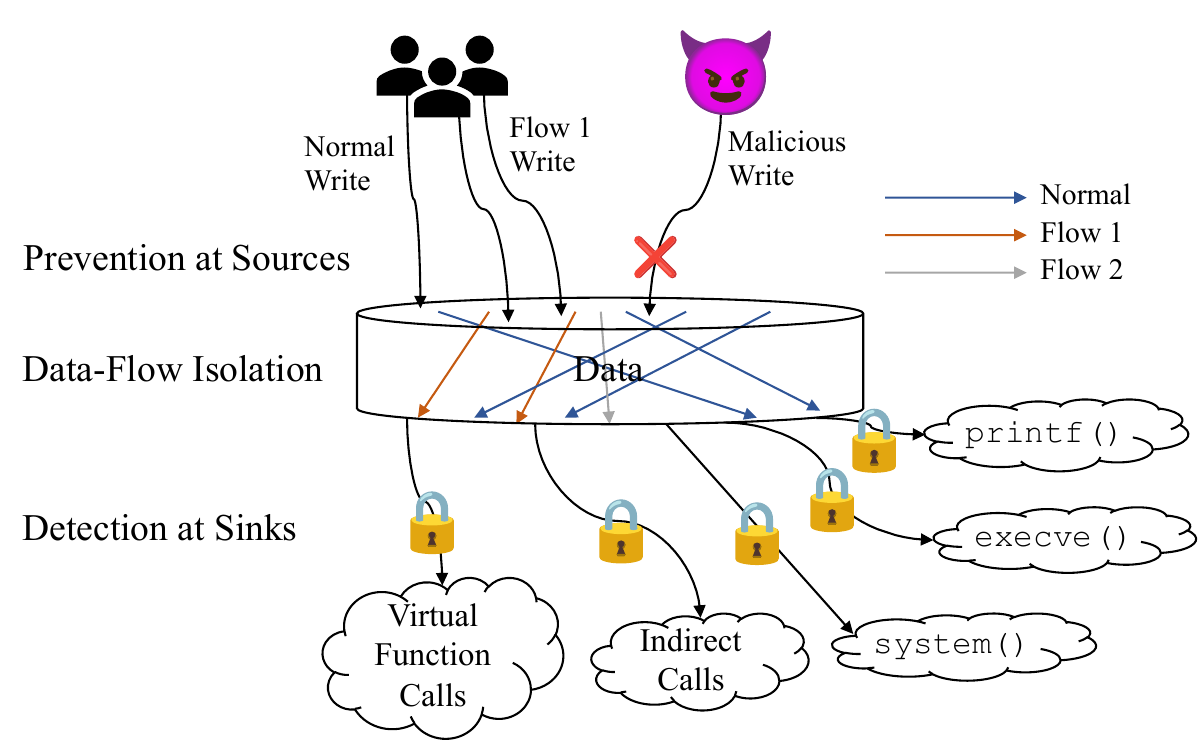}
    \vspace{-0.2cm}
    \caption{Data-Flow Integrity.
             Its mechanisms are divided into three groups,
             i.e., prevention at sources (e.g., Intel MPX, ARM MTE),
             data-flow isolation (e.g., HDFI, IMIX, Intel MPK, ARM DACR,
             Intel SMEP, Intel SMAP, ARM PAN, RISC-V SUM),
             and detection at sinks (e.g., ARM PA, \sysname).}
    \label{fig-dfi}
    \vspace{-0.4cm}
\end{figure}

\textbf{Data-Flow Integrity (DFI).}
DFI tries to ensure that the data-flow graph of user-mode programs~\cite{dfi, wit} or operating system kernels~\cite{kdfi} is not violated at runtime.
Data flows have their sources and sinks, and DFI mechanisms can be classified into three groups: (1) mechanisms that prevent maliciously memory writing at sources, (2) mechanisms that isolate sensitive data flows from normal data flows, and (3) mechanisms that detect attacks and validate the integrity of data at sinks, as shown in Figure~\ref{fig-dfi}.

\paragraph{Prevention at sources}
Intel MPX~\cite{intel-mpx} is a hardware extension designed to efficiently perform bound checks and prevent corresponding memory corruption attacks.
However, due to maintenance burdens and lack of industry uses, its support is removed.
ARM MTE~\cite{arm-mte} provides an ingenious mechanism of associating words with \textit{tags}, and can be used to build hardware-assisted address sanitizers~\cite{hwasan}.
However, it will take considerable hardware resources to implement and take mandatory time and memory overheads to store, set, and update the fine-grained tags.
Our prior experiments on Google Pixel 8 Pro (Tensor G3 processor) using Geekbench 6.2.0 show that ARM MTE introduces $\sim7\%$ execution-time overheads and $\sim3.125\%$ memory-usage overheads.

\paragraph{Data-flow isolation}
Some solutions provide DFI by isolating sensitive data like code pointers.
HDFI~\cite{hdfi} can provide strong data-flow isolation and security guarantees by associating each word (or several words) with one-bit tags.
However, because of these tags, it is complex and also requires considerable hardware resources to implement.
Besides, the actual hardware resource cost is not reported~\cite{hdfi}.
Similar to HDFI, IMIX~\cite{imix} is a lightweight page-grained data-flow isolation solution, which associates pages with one-bit tags.
It is similar to other isolation mechanisms and provides a coarse-grained version.
And, appropriate deployments of these mechanisms may require more manual effort.

Intel MPK~\cite{intel-sdm, libmpk} and ARM DACR~\cite{arm-dacr} both divide memory pages into several \textit{keys} or \textit{domains}, and provide handy control registers for users to adjust page permissions without context switching or invalidating TLBs.
They can be used to implement data-flow isolation schemes~\cite{erim}.

Intel SMAP-bit (Supervisor Mode Access Prevention)~\cite{intel-sdm}, preventing kernel-mode code from accessing user pages, is also a kind of isolation mechanisms that isolate kernel data from user data~\cite{seimi}.
Similarly, Intel SMEP-bit (Supervisor Mode Execution Prevention)~\cite{intel-sdm} prevents processors from executing user code in kernel mode, isolating kernel code from user code.
ARM PAN-bit (Privileged Access Never) has the same functionalities as Intel SMAP-bit~\cite{panic}.
RISC-V prohibits kernel mode from executing user code by default, and provides a SUM-bit (permit Supervisor User Memory access) having similar functionalities to Intel SMAP-bit.

\paragraph{Detection at sinks}
Solutions like ARM PA~\cite{arm-pa} and StackGuard~\cite{cowan1998stackguard} can detect corruption before sensitive data are used.
The hardware feature ARM PA enables defenses (e.g., PARTS~\cite{pacitup}) to validate the integrity of pointers at all sinks (i.e., pointer dereference points).
Specifically, pointers' Pointer Authentication Codes (PACs) are computed from pointers and contexts (i.e., modifiers) using cryptographic hash functions (e.g., QARMA~\cite{qarma}) at sources.
However, these cryptographic hash functions may increase the area and power consumption of processor cores as it will take more hardware resources~\cite{qarma}.
Attackers can also reuse existing pointers with PACs at any sensitive sink that uses the same modifier, so the security strength of ARM PA is equivalent to that of \sysname when the used length of modifiers equals the length of memory region keys.
In addition, ARM PA relies on the kernel to protect encryption keys, and thus is not suitable for systems without kernel-user separation.
Instead, \sysname also works at sinks, but provides a lightweight and practical pointee integrity mechanism, suitable for a wide range of systems.

\paragraph{Hiding-Based Defenses}
Randomization can hide secrets in random locations in memory space, and thus can protect secrets from corruption.
For example, ASLR is an efficient defense technique by hiding code and data, but is found vulnerable to information leak attacks or guessing-based attacks~\cite{aslr}.
Morpheus~\cite{morpheus} is a hardware randomization solution, which periodically re-randomizes and encrypts addresses at runtime by hardware.
It provides very strong security guarantees, but the complexity and overheads hinder its deployment.

EXecute-Only Memory~\cite{uxom} hides code and mitigates code reuse attacks by denying all executable data (i.e., code) being read.
Like DEP, it can be also implemented by configuring proper page permissions (e.g., only setting X-bit on RISC-V ISA).
However, it cannot protect other types of secrets.

%% file: concl.tex
\section{Conclusion}

In this paper, we have presented a lightweight hardware-software co-design solution \sysname to stop tainted data from being used at sensitive sinks.
The core of \sysname is a new set of instructions that only load data from read-only memory regions tagged with specific keys, which provide pointee integrity guarantees.
\sysname can be utilized by program hardening solutions to protect sensitive sinks with very low performance overheads and high security guarantees.
We have extended the RISC-V ISA and implemented an FPGA-based prototype of \sysname, and have demonstrated an important specific defense application of \sysname.
Evaluation results have shown that \sysname takes few extra hardware resources (\hwoverheads), and defense solutions based on \sysname introduce negligible runtime overheads (\overheads), while providing strong protection comparable to type-based CFI.

We believe that because of its simplicity and security guarantees, \sysname can be widely deployed to mitigate a large portion of memory corruption attacks in practice.